\documentstyle[aps,twocolumn,epsfig,floats]{revtex}

\def\Tr{{\bf Tr}\;}
\def\tr{{\bf tr}\;}

\def\ORDER#1{\hbox{${\cal O}(#1)$}}
\def\KET#1{\hbox{$|#1\rangle$}}

\def\BRACKET#1#2{\hbox{$\langle#1|#2\rangle$}}

\def\cross{\otimes}

\begin{document}

\title{Number Partitioning on a Quantum Computer}

\author{H. De Raedt$^1$, K. Michielsen$^1$, K. De Raedt$^2$,
and S. Miyashita$^3$}

\address{$^1$Institute for Theoretical Physics and
Materials Science Centre,\\
University of Groningen, Nijenborgh 4,\\
NL-9747 AG Groningen, The Netherlands\\
http://rugth30.phys.rug.nl/compphys\\
$^2$EMS, Vlasakker 21, B-2610 Wommelgem, Belgium\\
$^3$Department of Applied Physics, School of Engineering \\
University of Tokyo, Bunkyo-ku, Tokyo 113, Japan \\
E-mail: deraedt@phys.rug.nl, kristel@phys.rug.nl, miya@yuragi.t.u-tokyo.ac.jp\\
}

\date{DRAFT: \today}
\maketitle
\begin{abstract}
We present an algorithm to compute the number of solutions of the
(constrained) number partitioning problem.
A concrete implementation of the algorithm on an Ising-type quantum
computer is given.

\end{abstract}
\bigskip

\section{Introduction}

The discovery of quantum algorithms (QA's) that, when executed
on a quantum computer (QC), give significant speedup over
their classical counterparts \cite{SHORone,GROVERone}
has given strong impetus to recent developments
in the field of quantum computation.
In theory an ideal QC is a universal computer.
This means that for a given problem there exists
an algorithm to solve this problem on a QC.
The fact that a QC is a universal computer
does not tell us the algorithm itself, nor the
computational resources that are required to solve the problem.

In general it is not easy to construct algorithms for an ideal QC.
In particular, algorithms
that involve many conditional jumps (e.g. IF-THEN-ELSE statements)
are difficult to implement. %  on a physically realizable QC.
In essence this is because testing for a condition
requires a measurement during the execution of the program.
It is therefore of interest to see how a QC can solve problems
that a conventional computer solves by performing many conditional jumps.
The purpose of this paper is to present a new QA for
one such problem of combinatorial optimization:
The (constrained) number partitioning problem.

\section{Number Partitioning}

The number partitioning problem (NPP) is defined as follows
\cite{KARPone,GAREYzero,CORMENzero}: Does there exist a partitioning
of the set $A=\{a_1,\ldots,a_n\}$ of $n$ positive integers $a_j$
into two disjoint sets $A_1$ and $A_2=A-A_1$ such that
$\sum_{a_j\in A_1} a_j=\sum_{a_j\in A_2} a_j$ ?
The answer to this question is trivially no if the sum of all $a_j$,
$B\equiv\sum_{a_j\in A} a_j$, is odd.
More generally, the case of even or odd $B$ can be treated on the same footing
by asking if there exists a partition such that
$|\sum_{a_j\in A_1} a_j-\sum_{a_j\in A_2} a_j|\le\Delta$
where $\Delta=1$ ($0$) if $B$ is odd (even).
%In this paper we will use the latter formulation.

For certain applications there may be additional constraints
on the partitioning of the set $A$. A common one is to fix
the difference $C$ between the number of
elements in $A_1$ and $A_2$: $C\equiv\sum_{a_j\in A_1} 1-\sum_{a_j\in A_2} 1$.
For instance, if $C=0$ we ask if there is a partitioning
such that the number of elements in $A_1$ and $A_2$ are the same.

For a given instance of $A=\{a_1,\ldots,a_n\}$,
we may encode the whole problem using only $n\log_2B$ bits.
The NPP can be solved by dynamic programming, in
a time bounded by a low order polynomial in $nB$ \cite{GAREYzero}.
As $nB$ is not bounded by any polynomial of
the input size $n\log_2B$, the dynamic programming
algorithm does not solve the NPP with polynomial computational
resources\cite{GAREYzero}.

Number partitioning is one of Garey and Johnson's six basic NP-complete problems
\cite{GAREYzero}. In practice, a problem is NP-complete if its solution requires
computational resources that increase faster that any polynomial of the input
size. Number partitioning is a key problem in the theory of computational
complexity and has a number of important practical applications such as
job scheduling, task distribution on multiprocessor machines,
VLSI circuit design to name a few.

The computation time to solve a NPP depends
on the number of bits $b=\log_2 B$ needed to represent the integers $a_j$ and $B$.
Numerical simulations using random instances of $A$
show that the solution time grows exponentially with
$n$ for $n\ll b$ and polynomially for $n\gg b$
\cite{GENTone,GENTtwo,KORFone,MERTENSthree}.
For random instances $A$, the NPP can be mapped onto
a hard problem of statistical mechanics, namely that
of finding the ground state of an infinite-range Ising spin glass
\cite{MERTENSone,FERREIRAone,MERTENStwo}.
The transition from the computationally ``hard'' (exponential)
to ``easy'' (polynomial) has been related to the phase transition
in the statistical mechanical system \cite{MERTENSone,MERTENStwo}.

Although the transition between easy and hard problems
is important from conceptual point of view, it is good
to keep in mind that most real-life problems are of the easy type\cite{GAREYzero}.
For instance, if the $a_j$'s represent the weight of boxes that
are to be distributed over several trucks, it is highly unlikely
that the weight of these boxes will vary between say $1$kg and $2^{32}$kg,
or that it is important to know the weights of the boxes with a precision
of e.g. ten digits.

\section{Quantum Algorithm}

The potential power of a QC stems from the fact that a QC operates
on superpositions of states
\cite{FEYNMAN,LLOYDone,DIVINCENZOone,EKERTone,VEDRALone,BERMANzero,SHORtwo}.
The interference of these states allows exponentially
many computations to be done in parallel
\cite{FEYNMAN,LLOYDone,DIVINCENZOone,EKERTone,VEDRALone,BERMANzero,SHORtwo}.
A QA consists of a sequence of unitary transformations
that change the state of the QC
\cite{FEYNMAN,LLOYDone,DIVINCENZOone,EKERTone,VEDRALone,BERMANzero,SHORtwo}.
Therefore to solve a NPP on a QC, we first have to develop an algorithm that
does not contain conditional branches and
can be expressed entirely in terms of unitary operations.

A generic $n$-qubit QC can be modeled by
a collection of $n$ two-state systems, represented by
$n$ Pauli-spin matrices $\{\vec\sigma_1,\ldots,\vec\sigma_n\}$
\cite{FEYNMAN,LLOYDone,DIVINCENZOone,EKERTone,VEDRALone,BERMANzero,SHORtwo}.
The two eigenstates of $\sigma^z_j$ will be denoted by
$\KET{\uparrow}_j$ and $\KET{\downarrow}_j$,
corresponding to the states
$\KET{0}_j$ and $\KET{1}_j$ of the $j$-th qubit respectively.
The eigenvalues corresponding to
$\KET{\uparrow}_j$ and $\KET{\downarrow}_j$ are
$S_j=+1$ and $S_j=-1$. They can be
used to represent a partitioning of $A$:
We assign $a_j$ to $A_1$ ($A_2$) if $S_j=+1$ ($S_j=-1$).
If we can find a set $\{S_1,\ldots,S_n\}$
such that $\Delta - \sum_{j=1}^n a_j S_j=0$ we have found
one solution of the NPP.

It is known that the most simple class of spin
systems, i.e. those involving interactions of the Ising type only,
can be used to build universal QC's \cite{LLOYDone,BERMANzero,BERMANone}.
For our purposes it is, at this stage, sufficient to consider a system of
$n$ non-interacting Ising spins. The Hamiltonian is given by

\begin{equation}
\label{HAM}
H=\Delta-\sum_{j=1}^n a_j \sigma^z_j,
\end{equation}%
where the $a_j$'s represent external fields acting on the spins.
From (\ref{HAM}) it follows directly that
an eigenstate of $H$ with energy zero corresponds to
a solution of the NPP.
We will use Hamiltonian (\ref{HAM}) to define the time evolution of
the QC, i.e. the QA that solves NPP's.

The second key to the construction of the quantum algorithm
is the observation that the number of solutions $n_s$ of a NPP is given by

\begin{equation}
\label{Ns}
n_s\equiv\frac{1}{M} \sum_{m=0}^{M-1} \Tr e^{-2\pi i m H/M},
\end{equation}%
where $M\equiv B+\Delta+1$
and $\Tr U$ denotes the trace of the matrix $U$ \cite{COMMENT}.
Using the representation that diagonalizes the spin operators $\sigma^z_j$,
we find

\begin{eqnarray}
\label{Nsnul}
n_s&=&
\sum_{\{S_j=\pm1\}}
\frac{1}{M} \sum_{m=0}^{M-1}
\exp\left[ \frac{2\pi i m}{M} (\sum_{j=1}^n a_j S_j-\Delta) \right]
\nonumber\\
&=&
\sum_{\{S_j=\pm1\}}
\frac{1-\exp\left[ 2\pi i (\sum_{j=1}^n a_j S_j-\Delta) \right]}
{1-\exp\left[ 2\pi i (\sum_{j=1}^n a_j S_j-\Delta)/M \right]}
\nonumber\\
&=&
\sum_{\{S_j=\pm1\}} \delta_{\Delta,\sum_{j=1}^n a_j S_j}
.
\end{eqnarray}
As $|\Delta-\sum_{j=1}^n a_j S_j| < M$ for any choice of $\{S_j\}$,
the sum over $m$ in (2) will be zero unless
$\Delta-\sum_{j=1}^n a_j S_j=0$, in which case
the configuration $\{S_1,\ldots,S_n\}$ is
a solution of the NPP (note that there can
be exponentially many solutions, for instance in the case that all the $a_j$'s are equal).
Performing the sum over all spin configurations as indicated
in (\ref{Nsnul}), it follows immediately that $n_s$ is the number
of solutions of the NPP.
Note that (2) gives the number of
solutions of a NPP, which is more than just a yes or no answer to the question
if a partition of $A$ exists \cite{GAREYzero}.

Formally expression (\ref{Ns}) is the density of states at zero energy of the physical
system described by Hamiltonian (\ref{HAM}). Elsewhere we have shown
that for a large class of models $H$, the density of states can be calculated
efficiently on a QC \cite{HDR2}. The algorithm presented below, although
related to the one described in \cite{HDR2}, is specifically tuned to solve
NPP's.

The equivalence of (2) and the solution of the NPP can also be
demonstrated by explicit calculation of the trace over all spin configurations.
This is easy because the spins do not interact.
The result is

\begin{equation}
\label{Nss}
n_s=2^nM^{-1} \sum_{m=0}^{M-1}
e^{-2\pi i m\Delta/M} \prod_{j=1}^n \cos(2\pi m a_j/M).
\end{equation}%
For $\Delta=0$ and in the limit $M\rightarrow\infty$ we have
$n_s=2^n I_s$ where
\begin{equation}
I_s=\frac{1}{2\pi}\int_{0}^{2\pi} \cos (a_1 \theta)\ldots\cos (a_n \theta) d\theta.
\end{equation}%
The question whether $I_s=0$ or not is known to be equivalent to
the (non-)existence of a solution of a NPP\cite{GAREYzero,PLAISTEDone}.

If $n_s>0$ we can find a partitioning in the following manner.
Assume we already know the values of the first
$0<l-1<n$ spins. We make a guess for $S_{l}$ and compute
$n_s^{(l)}\equiv M^{-1} \sum_{m=0}^{M-1} \tr e^{-2\pi i m H/M}$
where the use of the symbol {\bf tr} instead of
{\bf Tr} indicates that in calculating the trace,
the values of the variables $S_{1},\ldots,S_{l}$ are fixed.
If $n_s^{(l)}>0$ our guess for $S_{l}$ was correct, if not
we reverse $S_{l}$. Then we increase $l$ by one and
repeat the procedure.

The algorithm outlined above is easily generalized to handle constraints.
Introducing another Hamiltonian

\begin{equation}
\label{Hprime}
H^\prime=C-\sum_{j=1}^n \sigma^z_j,
\end{equation}%
the number of solutions $n_s(C)$ of the constrained NPP is given by

\begin{equation}
\label{NsC}
n_s(C)\equiv\frac{1}{MK} \sum_{k=0}^{K-1}\sum_{m=0}^{M-1}
\Tr e^{-2\pi i m H/M} e^{-2\pi i k H^\prime/K},
\end{equation}%
where $K=n+|C|+1$.
Repeating the same steps as above we find
that the sum over $k$ yields zero unless
$C=\sum_{j=1}^N S_j=\sum_{a_j\in A_1} 1-\sum_{a_j\in A_2} 1$.
The expression corresponding to (\ref{Nss}) reads

\begin{eqnarray}
\label{NsCs}
n_s(C)=\frac{2^n}{MK} \sum_{k=0}^{K-1}\sum_{m=0}^{M-1}
&&e^{-2\pi i m\Delta/M - 2\pi i kC/K}
\nonumber\\
&&
\times \prod_{j=1}^n \cos\left(\frac{2\pi m a_j}{M}+\frac{2\pi k}{K}\right)
.
\end{eqnarray}
The procedure to find a partitioning itself
is the same as in the unconstrainted case.

The algorithms defined by (\ref{HAM}),(\ref{Ns}) and (\ref{Hprime}),(\ref{NsC}) solve NPP's
and constrained NPP's without recourse to dynamic programming.
This follows directly from explicit expressions (\ref{Nss})
and (\ref{NsCs}).
On a conventional computer the computation time required is
bounded by $nM$ (or $nMK$ for the constrained case).
Hence also these algorithms do not solve the (constrained) NPP
in polynomial time (space).
The conceptual difference between these algorithms and
the dynamic-programming approach is that the former
directly compute the number of solutions of the NPP whereas
the latter performs a search for a solution of the NPP.

As we now show, (\ref{Ns}) (or (\ref{NsC})) is a convenient
starting point to construct a QA that solves
the (constrained) NPP on a QC.
As will be clear from the discussion below, it will be
sufficient to concentrate on (\ref{Ns}), the algorithm for
(\ref{NsC}) is almost identical.

The first step is to introduce a ``number operator'' $X$
with eigenstates \KET{x}, $X\KET{x}=x\KET{x}$, $x=0,1,\ldots,M-1$.
We modify the Hamiltonian that governs the time-evolution of the
QC as follows:

\begin{equation}
\tilde H=\Delta X-\sum_{j=1}^n a_j \sigma^z_j X.
\end{equation}%
Calculating the trace in the basis that diagonalizes
$\sigma_1^z\ldots\sigma_n^z$ and $X$
we find that $n_s=M^{-1}\Tr e^{-2\pi i\tilde H/M}$.
Because $\tilde H$ is diagonal in this basis
$\Tr e^{-2\pi i\tilde H/M}$ is proportional
to one matrix element, namely

\begin{eqnarray}
\label{Nsss0}
\Tr &&e^{-2\pi i \tilde H/M}=
\nonumber\\
&&2^{n}M
\BRACKET{U_1\ldots U_n U_x}{e^{-2\pi i \tilde H/M}|U_1\ldots U_n U_x}
,
\end{eqnarray}%
where
$\KET{U_j}\equiv(\KET{\uparrow}_j+\KET{\downarrow}_j)/\sqrt{2}$
is the uniform superposition
of the spin up and down state of spin $j$, and
$\KET{U_x}\equiv(\KET{0}+\KET{1}+\ldots+\KET{M-1})/\sqrt{M}$
is the uniform superposition
of all the eigenstates of the number operator $X$.
To derive expression (\ref{Nsss0}) we made use of

\begin{equation}
e^{-ia \sigma_j^z}\KET{U_j}=\cos(a) \KET{U_j} -i\sin(a)\KET{\bar U_j},
\end{equation}%
and $\BRACKET{U_j}{\bar U_j}=0$,
where $\KET{\bar U_j}=(\KET{\uparrow}_j-\KET{\downarrow}_j)/\sqrt{2}$.

From (\ref{Ns}) it follows that

\begin{equation}
\label{Nsss}
n_s=2^{n}\BRACKET{U_1\ldots U_n U_x}{e^{-2\pi i \tilde H/M}|U_1\ldots U_n U_x}.
\end{equation}%
As a QC can compute $e^{-itH}\KET{\psi}$ with one operation (for arbitrary input
\KET{\psi}) \cite{FEYNMAN},
(\ref{Nsss}) shows that once the QC is in the state of uniform superposition
\KET{U_1 \ldots U_N U_x}, one time-evolution step of the QC will solve the NPP.

The initial state \KET{\uparrow,\ldots,\uparrow,x=0} can be transformed
into the state of uniform superposition \KET{U_1 \ldots U_N U_x}
by the standard procedure \cite{EKERTone,VEDRALone}:
The states $\KET{U}_j$ can be obtained from the initial state $\KET{\uparrow}_j$
by a rotation of the spin $j$ about the $y$-axis, i.e.
$\KET{U}_j=e^{-i\pi\sigma^y_j/4}\KET{\uparrow}_j$ for $j=1,\ldots,n$.
On an Ising-type QC the states \KET{x} can be implemented by
adding new two-state systems.
We denote the corresponding Pauli-spin operators and eigenvalues
by $\vec\mu_p$ and $s_p$ respectively.
We use these spins to represent
$x=\sum_{l=1}^p 2^{l-2} (1-s_l)$ in binary form.
As $0\le x< M$ the number of spins $p$
required to represent $x$ is the smallest integer $p$ for
which $M\le2^p$.
Using this binary representation for \KET{x}, the uniform superposition
\KET{U_x} can be obtained by $p$ rotations of the initial state:

\begin{equation}
\KET{U_x}=e^{-i\pi\mu^y_1/4}\KET{\uparrow}_1\cross\ldots
\cross e^{-i\pi\mu^y_p/4}\KET{\uparrow}_p,
\end{equation}%
where $\cross$ denotes the direct product operation.
The system now comprises $n+p$ spins and its Hamiltonian reads

\begin{equation}
\label{Hcal}
{\cal H}=
-\sum_{l=1}^p \sum_{j=1}^n J_{j,l}^{\phantom{z}} \sigma^z_j\mu^z_l
-\sum_{j=1}^n b_{j} \sigma^z_j
-\sum_{l=1}^p c_l\mu^z_l
+d,
\end{equation}%
where $J_{j,l}=-a_j2^{l-2}$,
$b_j=a_j(2^{p}-1)/2$,
$c_l=\Delta 2^{l-2}$,
and
$d=\Delta (2^{p}-1)/2$.

The complete QA for computing $n_s$, i.e. for solving
NPP's, can be summarized as follows:
The initial state of the QC (all spins up by convention) is transformed
into the state of uniform superposition. This takes $n+p$ one-qubit
operations. Next the QC makes one time-evolution step $\exp(-i\pi {\cal H}/2^{p-1})$.
The matrix element in (\ref{Nsss}) is obtained by applying the inverse of the
$n+p$ rotations, followed
by a projection onto the initial state. Clearly the total number
of QC operations is only $2n+2p+1$ while the amount of memory used
is \ORDER{\log_2 M + \log_2 n}.

The constrained NPP can be solved in the same way: Add
qubits to represent the variable $k$ in (\ref{NsC}) and repeat the steps
that lead to (\ref{Nsss}).
Note that once the uniform superposition has been prepared,
the QA also solves the constrained NPP with one time-evolution step.

\section{Implementation on a Quantum Computer Emulator}

For the purpose of demonstration we have implemented the QA
that solves the unconstrained NPP on a Quantum Computer Emulator (QCE),
a software tool for simulating physical models of QC's \cite{QCEzero}.
A subtle point thereby is that (\ref{Nsss}) is
not directly measurable because $e^{-2\pi i \tilde H/M}$ is
not a physical observable.
However it is not difficult to express $n_s$ in terms
of an expectation value of a physical observable.

Let us write the number of solutions (7) as
$n_s=2^{n}\BRACKET{0}{\Phi}$ where
$\KET{\Phi}=U^{-1}e^{-i\pi {\cal H}/2^{p-1}}U\KET{0}$
and $U\equiv e^{-i\pi\sigma^y_1/4}\ldots e^{-i\pi\sigma^y_n/4}
e^{-i\pi\mu^y_1/4}\ldots e^{-i\pi\mu^y_p/4}$.
Our aim is to replace the projection onto the initial state \KET{0},
a shorthand notation for the state with all spins up, by the measurement
of some observable.
This can be accomplished by introducing another spin $\vec\kappa$, initially
in the state of spin up, to the system
and flip this spin if the other $n+p$ are all up, i.e. by performing
an AND operation on the $n+p$ spins.
With $V$ denoting the unitary transformation that performs this AND operation,
we have in the language of qubits instead of spins

\begin{eqnarray}
\label{PSI}
\KET{\Psi}&\equiv& V U^{-1}e^{-i\pi {\cal H}/2^{p-1}}U\KET{0}\cross\KET{0}_\kappa
\nonumber\\
         &=&V [2^{-n} n_s\KET{0}\cross\KET{0}_\kappa+(\ldots)\cross\KET{0}_\kappa]
\nonumber\\
         &=&2^{-n} n_s\KET{0}\cross\KET{1}_\kappa+(\ldots)\cross\KET{0}_\kappa,
\end{eqnarray}%
where \KET{\Psi} is an element of the direct product of the Hilbert spaces
spanned by the $n+p$ spins and the auxillary spin $\vec\kappa$.
We use the abbreviation $(\ldots)$ to represent the sum of all
states of the $n+p$ spins that have at least one spin down.
From (\ref{PSI}) it immediately follows that

\begin{equation}
\label{Nsfinal}
n_s=2^{n}\BRACKET{\Psi}{(1-\kappa^z)/2|\Psi}^{1/2}.
\end{equation}

It is well-known how to implement the AND operation on a QC \cite{BARENCOone}.
In our practical implementation \cite{WWW}, we have choosen to use
a three-bit network, the Toffoli-gate,
as a building block for realizing the AND operation on the
$n+p$ qubits \cite{BARENCOone}.
By adding extra work qubits the complete network
requires of the order of $\log_2(n+p)$ steps and $n+p$ extra qubits
to perform the AND operation. Clearly this does not change
the polynomial time and space character of the QA that
solves NPP's. A block diagram of the complete quantum program is shown
in Fig.1.
We have implemented the QA on a 15-qubit QC and used it
to solve the NPP's $A=\{1,2,3,4\}$, $A=\{1,1,1,4\}$ and $A=\{2,2,2,4\}$
(these examples are included in the software distribution \cite{WWW}).
In the final state the expectation values of the 15-th qubit
are $0.015625$, $0.00390625$, and $0$ respectively.
The corresponding number of solutions is $n_s=2$, $n_s=1$ and $n_s=0$.
Clearly the demonstration program correctly solves NPP problems.
%because we use a conventional computer to emulate the QC,
%it does not solve NPP with the efficiency of a genuine QC.

\section{Alternative implementation}

The implementation described above has the same logical
structure as other QA's \cite{SHORone,GROVERone,SHORtwo}:
Prepare the QC in a state of uniform superposition,
perform some unitary transformation to encode information
and then apply a filter to extract the answer.
We now show that there is another QA that solves the NPP
but does not fit into this general scheme in that the first
step is missing.

Consider the time-dependent $n$-spin correlation function

\begin{eqnarray}
\label{Ct}
%C(t)&=&\frac{\Tr e^{iH_xt}{\sigma^z_1\ldots\sigma^z_n e^{-iH_xt}\sigma^z_1\ldots\sigma^z_n}{\Tr\hbox{\bf{1}}}
%\nonumber\\
C(t)&=&\BRACKET{\Phi}{e^{iH_xt}\sigma^z_1\ldots\sigma^z_n e^{-iH_xt}\sigma^z_1\ldots\sigma^z_n|\Phi},
\end{eqnarray}%
where $H_x=-\sum_{j=1}^n a_j \sigma^x_j/2$.
The state \KET{\Phi} can be any $n$-spin state that is an eigenstate of
$\sigma^z_1\ldots\sigma^z_n$, e.g. the state with all spins up.
As the $\sigma^z_j$'s are unitary operators, it is a simple matter to write down
a QA that computes $C(t)$ on a QC.
Obviously $C(t)$ is a physically observable quantity
but it may require a rather complicated experimental setup to
measure this $n$-spin correlation function.

Substituting into (\ref{Ct}) the equation of motion for each spin, i.e.
$e^{iH_xt}\sigma^z_j e^{-iH_xt}=\sigma^z_j\cos(a_jt)-\sigma^y_j\sin(a_jt)$,
we find

\begin{eqnarray}
%C(t)&=&\frac{\Tr \prod_{j=1}^n[\hbox{\bf1}\cos(a_jt)-i\sigma^x_j\sin(a_jt)]}}{\Tr\hbox{\bf{1}}}
C(t)&=&\BRACKET{\Phi}{\prod_{j=1}^n[\hbox{\bf1}\cos(a_jt)-i\sigma^x_j\sin(a_jt)]|\Phi}
\nonumber\\
&=&\prod_{j=1}^n\cos(a_kt)
.
\end{eqnarray}%
The Fourier transform of $C(t)$ at zero frequency is directly
proportional to $I_s$ and hence to $n_s$:

\begin{eqnarray}
\label{Somega}
S(\omega)&=&\int_{-\infty}^{\infty} e^{i\omega t}C(t)dt=
2^{-n} n_s\delta(\omega) + R(\omega)
,
\end{eqnarray}%
where the regular part $R(\omega)$ is zero at $\omega=0$.
From (\ref{Somega}) it is clear that the NPP has a solution
if $S(\omega)$ shows a peak at zero frequency.
Detection of the central peak in the dynamic
correlation function $S(\omega)$ may require a very long observation time $T$.
To distinghuish between $n_s=0$ and $n_s=1$ the observation time $T$
must be larger than $2^n\pi$.

\section{Discussion}

The essence of the algorithm proposed above is that it expresses
$n_s$ in terms of the density of states of a physical system (Ising spins in our example).
Clearly this QA certainly has its weakness if $n_s$ is close to zero and $n$ is large.
Of the order of $2^n$ measurements of $\kappa_z$ are required
to distinguish between $n_s=0$, and $n_s =1$.
This is tantamount to random sampling.
By formulating, as we did, the outcome of the
calculation in terms of (an average of) physical observables (see (\ref{Nsfinal}) or (\ref{Somega})
instead of a (collapsed) state, this problem of efficiency is difficult to overlook
\cite{alsoshor}.

NP-completeness of the NPP refers to non-polynomial
behavior as a function of $nB$ ($B$ being the input size).
Our QA is polynomial in this respect.
However, in the hard case ($b\gg n$) and for large $n$, to obtain a yes-or-no answer
tremendous precision is required.
In the absence of any information of the $a_j$'s other than that they
are positive integers, the range of $n_s$ extends from zero to
${n}\choose{n/2}$.
Any algorithm that computes $n_s$ (on a conventional
or QC) should be able to cover this range (otherwise it can
never return the correct $n_s$). This implies that the whole computation
has to be done with at least the same (high) precision.

As the NPP example shows, a realistic assessment of the potential power of a QA
should include a quantitative estimate of the precision and other
computational resources (e.g. energy) that are required to obtain the correct answer.
For our QA an estimate of the required precision follows from (\ref{Nsfinal}).
Note that the alternative implementation yields a similar, physically
equivalent, estimate for the observation time $T$.

The range of numbers a physically realizable QC will be able to handle
is directly related to the energy range in which the QC operates
($-B$ to $+B$ in the NPP case).
Although not a problem of principle, the physics of
QC hardware will definitely impose some constraints on the
range of numbers.

There are two other, potentially large, numbers involved in an NPP problem:
The number of states $N_s = 2^n$ and the number of computational units $N_{cu}$ (of microscopic size)
in the physical realization of the QC.
There are two cases to consider.
1) Theoretical (computer science): We have to examine the worst case.
Then $N_s \gg N_{cu}$ so that our NPP algorithm has little merit.
2) In practice: In numerical experiments \cite{GENTone,GENTtwo,KORFone,MERTENSthree}
$n\le32$ ($N_s < 2^{32}$)
whereas for instance in NMR QC experiments $N_{cu}\approx 10^{18} \gg N_s$\cite{chuang}.
%Assuming it is physically possible to couple the spins to e.g. photons,
Using a sufficiently large number of computational units and efficient detectors
it should be possible to distinguish between $n_s=0$ and $n_s=1$.
Assuming that clever engineering can produce spin systems such as (\ref{Hcal}), our
QA might be used to demonstrate that a physical QC can solve a non-trivial problem.

This work is partially supported by the Dutch `Stichting Nationale Computer Faciliteiten'
(NCF).

\begin{figure}[ht]
\setlength{\unitlength}{1cm}
\begin{picture}(9.0,6.0)
\put(0,0.25){\epsfig{file=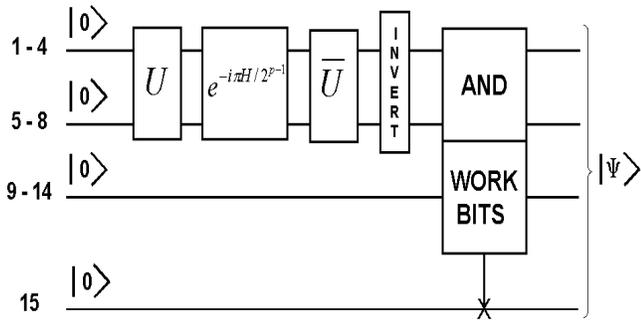,width=8.5cm,height=5.0cm}}
\end{picture}
\caption{%
Block diagram of the quantum algorithm that solves the number
partitioning problem in polynomial time and space. In this example
the first $n=4$ qubits are used to represent the integers to be
partitioned. The $p=4$ qubits 5 to 8 are used to determine the number
of solutions of the number partitioning problem. The remaining 7 qubits
are used to relate $n_s$ to a physically measurable quantity: The
expectation value of the 15-th qubit.
The unitary transformation $U$ prepares the uniform superposition
of the first 8 qubits, $\bar U$ is the inverse of $U$, and
the combination of INVERT and AND gates
sets the 15-th qubit to one if and only if the first eight qubits are all one.
}
\label{fig:fig1}
\end{figure}


\begin{thebibliography}{99}
\bibitem{COMMENT}
This value of $M$ is not optimal and can be reduced by a factor of about two.
However this is irrelevant for what follows and we therefore omit
the discussion of this technical point.
\end{thebibliography}
\end{document}